\documentclass[3p, twocolumn]{elsarticle}   

\journal{arxiv.org}

\usepackage{mathtools}
\usepackage[english]{babel} 
\usepackage{lipsum} 
\usepackage{xcolor}
\usepackage[export]{adjustbox}

\usepackage{graphicx}
\usepackage{dcolumn}
\usepackage{bm}
\usepackage{epstopdf}
\usepackage{color}
\usepackage{array,tabularx,longtable,multirow,xfrac}
\usepackage{placeins}

\DeclareRobustCommand{\citen}[1]{%
	\begingroup
	\romannumeral-`\x 
	\setcitestyle{numbers}%
	\cite{#1}%
	\endgroup   
}

\begin{document}

\begin{frontmatter}

\title{Structure of amorphous Cu$_2$GeTe$_3$ and a model for its fast phase-change mechanism}  
%
\author{Jens R. Stellhorn*$^,$\fnref{myfootnote}  }   
\ead{stellhorn@kumamoto-u.ac.jp}
\author{Benedict Paulus, Shinya Hosokawa}  

\address{%
	Department of Physics, Kumamoto University, Kumamoto 860-8555, Japan
}%
\address{%
	Department of Chemistry, Philipps University of Marburg, 35032 Marburg, Germany
}
\fntext[myfootnote]{Present affiliation: Deutsches Elektronen-Synchrotron DESY, 22603 Hamburg, Germany}
\cortext[mycorrespondingauthor]{Corresponding author}

\author{Wolf-Christian Pilgrim}%
\address{%
	Department of Chemistry, Philipps University of Marburg, 35032 Marburg, Germany
}%

\author{Nathalie Boudet, Nils Blanc}
\address{%
	Universit\'e Grenoble Alpes, CNRS, Grenoble INP, Institut N\'eel, 38000 Grenoble, France  
}%

\author{Yuji Sutou}
\address{%
	Department of Materials Science, Tohoku University, 
	Sendai 980-8579, Japan
}%


%
%
%
%

\begin{abstract}
The structure of amorphous Cu$_2$GeTe$_3$ is investigated by a combination of anomalous x-ray scattering and extended x-ray absorption fine structure experiments. The experimental data are analyzed with a reverse Monte Carlo modeling procedure, and interpreted in terms of the short- and intermediate-range order.
Based on this information, a model for the phase transition in Cu$_2$GeTe$_3$ is proposed, in which atoms move toward the center of the 6-fold rings of the crystal structure, leading to the formation of wrong bonds and a broader distribution of ring structures, but also preserving some structural motifs of the crystal. 
\end{abstract}


\end{frontmatter}


\section{Introduction}
Phase-change materials (PCM) utilize the reversible change from an amorphous phase to a crystalline  phase  to encode binary data.\cite{wuttig-nature2007, OvshinskyPRL}
The readability of the stored data is guaranteed by the pronounced differences in the electrical and/or optical properties of both phases. 
Cu$_2$GeTe$_3$ (CGT) is a new PCM which is expected to be used for a next generation of (non-volatile) data storage devices. \cite{Sutou2012,Saito2013}
The CGT crystalline film was found to be amorphized by laser irradiation with a lower power and shorter pulse width than currently employed GeSbTe alloys, which are essential properties to achieve rapid data recording and low power consumption in PCMs. \cite{Yamada1991, wuttig-nature2007, Saito2013}
In contrast to widely studied PCM systems like GeSbTe, amorphous CGT is denser than the crystal,  and the phase transition takes place to a tetrahedrally bonded crystal, a very different geometry compared to the octahedrally bonded cubic structures adopted by GeSbTe systems. \cite{Saito2013,Saito2014} 
A result of this peculiar structure is  a negative optical contrast, i.e.\ the reflectivity of the crystalline phase is lower than that of the amorphous phase,\cite{Saito2013} contrary to GeSbTe with a positive optical contrast.\cite{Yamada1991,shportko2008}

The structure of the crystal phase of CGT consists of a three-dimensional arrangement of slightly distorted corner-sharing CuTe$_4$ and GeTe$_4$ tetrahedra, with the space group $Imm2$. \cite{DelgadoGCT} (A schematic view is illustrated in the supplemental information.)
Concerning the structure of the amorphous phase, on the other hand, the state of research is inconsistent: 
different average coordination numbers have been reported by x-ray diffraction in combination with  x-ray absorption fine structure (XAFS) measurements,\cite{JovariGCT} by XAFS investigations alone \cite{Kamimura2016} and by \textit{ab-initio} molecular dynamics simulations~(AIMD).\cite{Skelton2013, Chen2015} The experimental results so far indicate that all atoms are roughly fourfold coordinated, which would constitute a large similarity to the tetrahedral crystal structure. AIMD simulations find much larger coordination numbers of Ge and Cu, with values exceeding 6 for Cu and  about 4.5 for Ge. 
Agreement is reached only on two points, namely the unusually large average coordination numbers for Cu and Te atoms, and  on the existence of a significant number of homopolar bonds for Cu-Cu and  Te-Te   pairs (so called ``wrong'' bonds, which do not exist in the crystalline phase).

Apart from the investigation of nearest neighbor arrangements, ring statistics calculations offer the possibility to characterize the topological connectivity of network structures. 
For GeSbTe, ring structures have been investigated both experimentally \cite{KoharaAPL} and theoretically by DFT simulations.\cite{AkolaJones, akola2009, HegedusNM} 
In general, the fast phase change ability of GeSbTe was attributed to a strong preference of (alternating) even-fold rings, facilitating the phase transition to the crystal with a similar ring structure. 
For CGT,  so far there are only theoretical investigations available.
\cite{Skelton2013,Chen2015} The reported ring structures are strikingly different from the known features of GeSbTe, especially concerning a large contribution of 3-fold rings. However, an experimental support for this finding is still missing.

The structural description also needs to be explained in the larger context of the transition from the crystal to the amorphous phase. 
Again, for GeSbTe, such models already exist and have been controversially discussed, e.g.\ the (modified) ``umbrella-flip'' model\cite{KolobovNatMat, hosokawa-GST} or the ring statistics analogy model\cite{KoharaAPL}. To build a suitable model for this process in CGT, detailed structural information on the short- and intermediate-range order of the amorphous phase are required. A powerful method to extract this kind of information is anomalous x-ray scattering (AXS). 
The aim of this article is thus to propose such a phase-change model for CGT,  based on a combination of anomalous x-ray scattering and extended XAFS experiments, analyzing the datasets with a reverse Monte Carlo (RMC) modeling procedure.

\section{Experimental}

The amorphous CGT sample was prepared by radio-frequency sputtering deposition from GeTe and CuTe alloy targets on SiO$_2$ (20~nm)/Si (0.7~mm) substrates. Details on the sample preparation are outlined in refs.~\citen{Saito2013} and~\citen{Saito2014}. We note that the as-deposited phase exhibits almost identical properties compared to the melt-quenched film that would be generated in a phase-change memory device.\cite{Saito2013} 

The AXS experiment were performed at the beamline BM02 of the European Synchrotron Radiation Facility (ESRF). 
AXS utilizes the anomalous variation of the atomic form factor $f$ of a specific element  in the vicinity of an x-ray absorption edge.\cite{Waseda1984} The experimentally accessible information are the differential structure factors $\Delta_kS(Q)$:

\begin{small}
\begin{align*}
\Delta_k S(Q) = \frac{ \Delta_k \left[C\cdot I(Q,E_{1},E_{2})\right] - \Delta_k \left[ \langle f^2\rangle - \langle f\rangle^2  \right]}{ \Delta_k \left[ \langle f\rangle^2 \right]} ,
\label{eq_theorie19}
\end{align*} 
\end{small}
which are calculated from the difference ($\Delta_k$) of two scattering experiments with intensities $I(Q,E)$ conducted at energies $E_1$ and $E_2$ close to the absorption edge. $C$ denotes the normalization factor.  The $\Delta_kS(Q)$ functions contain structural information specifically related to the element $k$. 
The relative increase of this information can be illustrated by the AXS weighting factors $w_{ij}$ for the partial contributions of all elements $i,j$: 
\begin{equation*}
\Delta_k S(Q) = \sum_{i,j} \Delta_k w_{ij}(Q)\cdot S_{ij}(Q).
\label{eq_theorie20}
\end{equation*}
They are illustrated for CGT in Table~\ref{tab:wij}. 
Note that the $w_{ij}$ have a small $Q$ dependence, and are given here exemplary for  $Q=1.9$~\AA$^{-1}$, i.e.\ at the first $S(Q)$ maximum position.
Incident energies for the measurements were selected 20 and 200~eV below the Cu and Ge $K$ edge, as well as 30 and 300~eV below the Te $K$ edge, respectively.
The experiments were performed in transmission geometry using a container cell with 7~$\mu$m Kapton windows and appropriate thicknesses for each investigated energy region. The data were corrected for absorption effects and Compton scattering, and normalized using the Krogh-Moe-Norman method.\cite{KroghMoe,norman}.
Further details on the theoretical and experimental background of AXS can be found elsewhere.\cite{Waseda1984, HosokawaPRB, stellhorn-zpc, HosokawaZPC, HosokawaEPJST} 

\begin{figure}
	\begin{center}
		\includegraphics[width=0.95\linewidth]{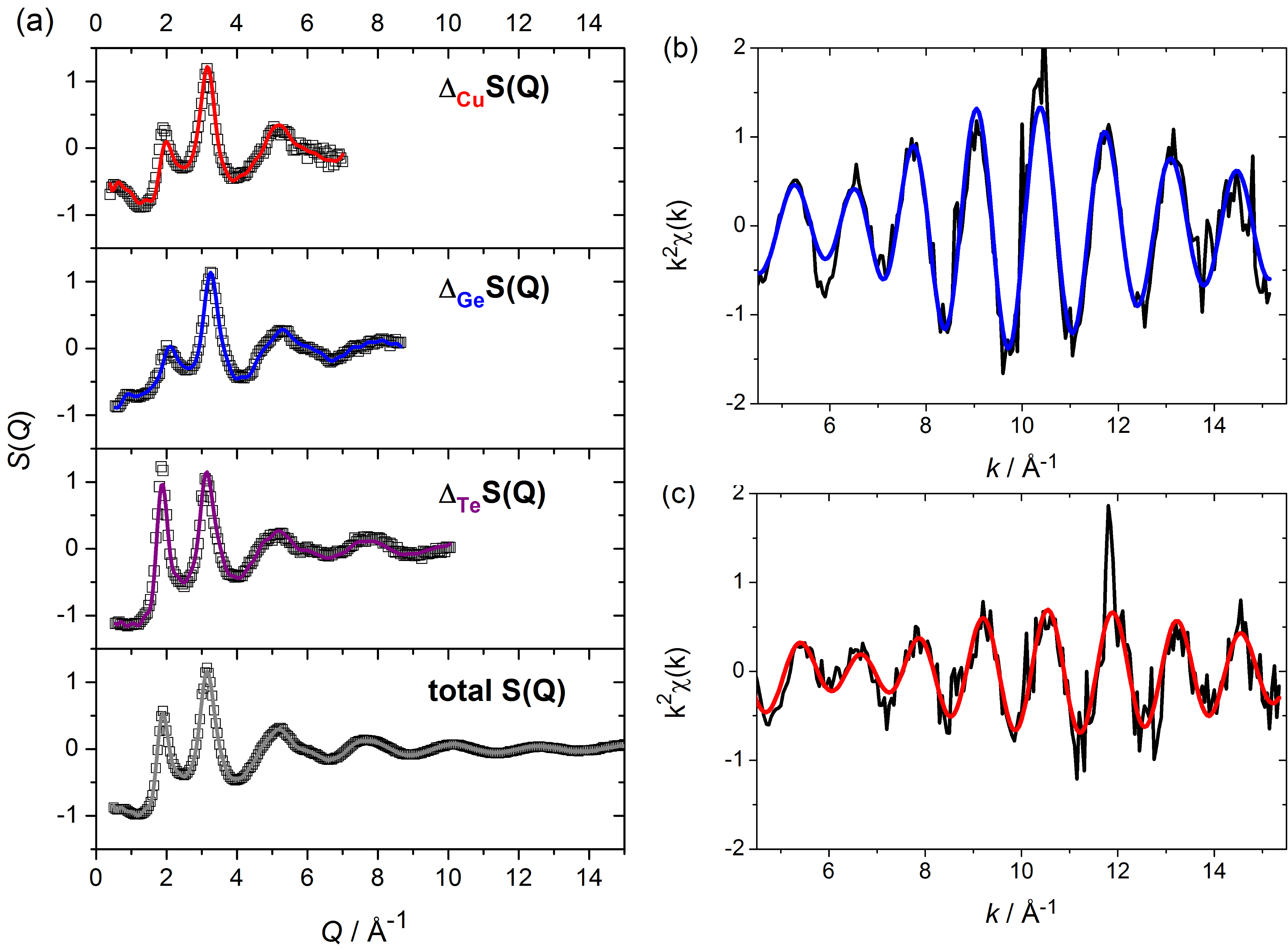}   
		\caption{Experimental data. (a) AXS, (b) Ge XAFS, (c) Cu XAFS. Black squares and black lines are experimental data. RMC fits are displayed as colored lines, with data acquired near the absorption edges of 
			Te (purple),
			Ge (blue) and 
			Cu (red),
			the total $S(Q)$ is shown in grey.
		}
		\label{fig:Exp_data}
	\end{center}
\end{figure}

The XAFS experiments were conducted at BL12C of the Photon Factory in the High Energy Accelerator Research Organization (KEK-PF), in fluorescence mode.  
The incident x-ray intensity was measured using an ion chamber, and the fluorescent x-ray intensity from the sample was detected using a 19-channel pure Ge solid state detector. XAFS functions were determined near the $K$ edges of Cu and Ge. 
Both AXS and XAFS data are displayed in Fig.~\ref{fig:Exp_data} with black symbols/lines. 

In the reverse Monte Carlo procedure, the real sample is modeled by an ensemble of atoms as hard spheres in a simulation box. In each simulation step, individual atoms are moved randomly to minimize the difference between experimental structure factors and those computed from the simulated configuration using a Metropolis algorithm. 
We employed the RMC\_POT package\cite{Gereben, Gereben2012} for our simulations.  
An input configuration of 10,000 atoms with proper stoichiometry with an initial random distribution of the atoms in a box corresponding to the number density of  0.0385~\AA$^{-3}$ was used. 
Minimal interatomic distances for the individual correlations Cu-Cu,   Cu-Ge,  Cu-Te   Ge-Ge, Ge-Te,   and Te-Te were set to 2.45,  2.35,  2.35,  2.35,  2.35  and 2.45~\AA, respectively. These distances were chosen near the values for the respective sums of the covalent radii,\cite{pyykko-covalent-radii}  and were adjusted to fit the first coordination shells adequately.

Two different conditions were explored with different sets of input data. First, in order to compare our findings with existing results, we included only the total structure factor $S(Q)$ (obtained at 300~eV below the Te $K$ absorption edge energy) and the two XAFS datasets in a RMC simulation, and excluded the formation of Cu-Ge bonds in the simulation by increasing the Cu-Ge minimal interatomic distance to 3.1~\AA\ (``limited approach''). The results were expected to be comparable to previously published data by J\'{o}v\'{a}ri \textit{et al.}~\cite{JovariGCT}, the only difference in the procedure being that no Te XAFS data were included in our model.
Secondly, the AXS data were included as well, i.e.\ the individual $\Delta_kS(Q)$'s for each element. This approach therefore illustrates the effect of the AXS data. Furthermore, the formation of Cu-Ge bonds was not excluded (``present model''). This kind of model reflects findings of the theoretical studies, in which no specific restriction to the possible types of bonds are found, i.e.\ Cu-Ge bonds are present.\cite{Skelton2013, Chen2015} 
The resulting RMC fits for this model are illustrated in Fig.~\ref{fig:Exp_data} with colored lines. 

To evaluate the significance of the Cu-Ge bonds, a different RMC run was performed with all of the experimental datasets, but excluding the Cu-Ge bond. For this, the minimal interatomic distance of the Cu-Ge pair was raised to 3.1~\AA. 
Relative to the simulations including the Cu-Ge bond, this leads to an increase in the goodness-of-fit values $R_w$ for the EXAFS dataset of Ge (by 8.4\%) and for the $\Delta_{\rm Ge}S(Q)$ function (by 7.2\%), and to a smaller amount for the  EXAFS dataset of Cu (by 2.1\%). This indicates the presence of Cu-Ge bonds in the material.

\setlength{\tabcolsep}{2pt}
\begin{table}
	\caption{Weighting factors $w_{ij}$ in CGT at 1.9~\AA$^{-1}$ near the first $S(Q)$ maximum, in percent.}
	\label{tab:wij}
	\begin{center}
		\begin{tabular}{l|cccccc}  
			\hline	
			& \small	Ge-Ge	&\small	Ge-Cu	&\small	Ge-Te	&\small	Cu-Cu	&\small	Cu-Te	&\small	Te-Te	\\
			\hline	
			$S(Q)$			&	1.9		&	7.0		&	16.8	&	6.5		&	30.9	&	36.9	\\
			$\Delta_{\rm Ge}S(Q)$	&	11.5	&	24.3	&	68.3	&	-1.7	&	-4.0	&	1.6	 \\
			$\Delta_{\rm Cu}S(Q)$   &	0.2		&	13.1	&	0.9		&	20.3	&	65.6	&	-0.1	\\
			$\Delta_{\rm Te}S(Q)$ 	&	0.0		&	0.0		&	14.0	&	0.0		&	25.8	&	60.1	\\
			%
			\hline
		\end{tabular}
	\end{center}
\end{table}

\section{Results}

\begin{figure}
	\begin{center}             
		\includegraphics[width=0.75\linewidth]{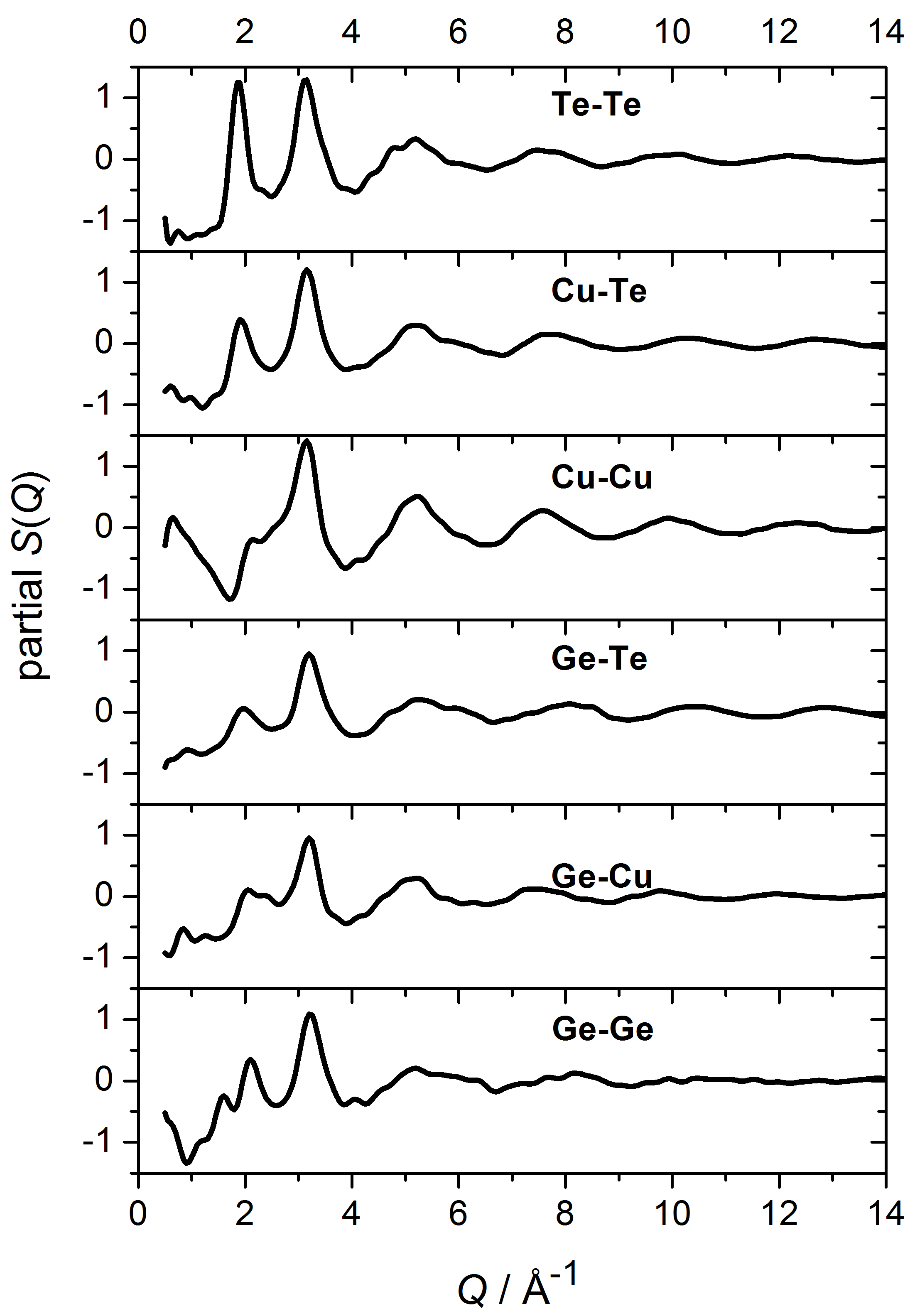} 
		\caption{RMC results for the partial structure factors $S_{ij}(Q)$ in the present model. }
		\label{fig:pfq_pfq}
	\end{center}
\end{figure}

\begin{figure}
	\begin{center}             
		\includegraphics[width=0.75\linewidth]{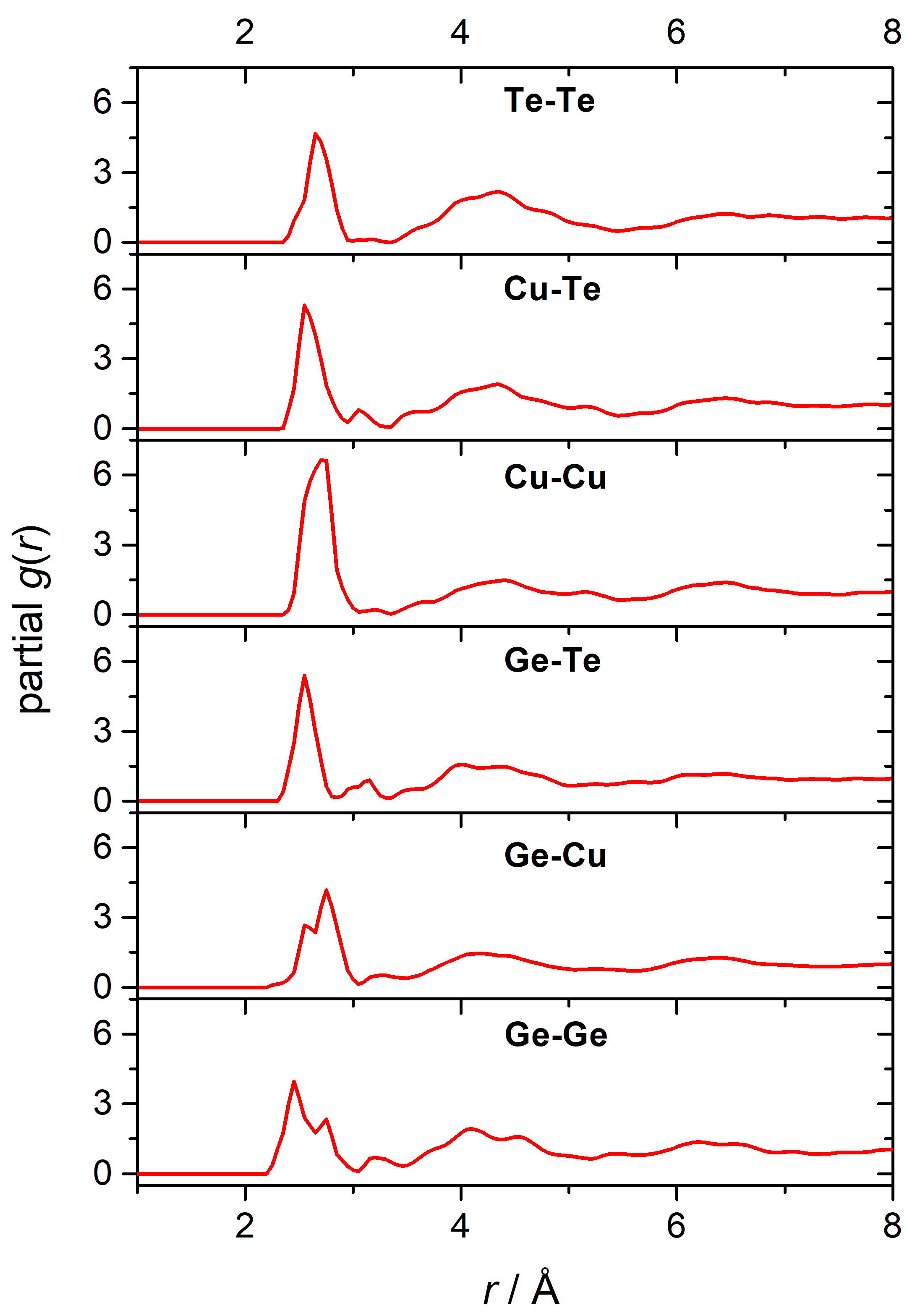}
		\caption{RMC results for the partial pair correlations functions $g_{ij}(r)$ in the present model. }
		\label{fig:pfq_ppcf}
	\end{center}
\end{figure}

From the RMC generated models, the 6 independent correlations of element pairs in CGT are calculated.  
Figures~\ref{fig:pfq_ppcf} and \ref{fig:pfq_pfq} give an overview of all partial structure factors $S_{ij}(Q)$ and pair correlation functions  $g_{ij}(r)$ obtained from the RMC simulation for the present model. Average bond lengths extracted from the pair correlations are listed in Table~\ref{tab:Disttable}, in comparison with data from two other studies and the corresponding values for the CGT crystal. The bond lengths are averaged over all correlations of the respective element. 
In general, bond lengths become slightly larger in the amorphous state compared to the crystal. The largest differences between different approaches is observed for Cu-related bonds: contrary to ref.~\cite{JovariGCT}, we find that the distances become somewhat larger than in the crystal, but the elongation is less compared to AIMD results \cite{Skelton2013}.  
Note that the precision of RMC with respect to interatomic distances in this approach  is around $\pm0.05$~\AA.

The partial and total coordination numbers are tabulated in Table~\ref{tab:CNtable}. Cut-off distances for the calculation of the coordination numbers were set to the first minimum in the pair correlation functions around 3.0~\AA.

\begin{table}
	\caption{Average bond lengths in~\AA\ for each element, in comparison with other studies. } 
	\label{tab:Disttable}
	\begin{center}
		\begin{tabular}{c|cccc}  
			\hline 
			species 	&	present	&	RMC\cite{JovariGCT}	& AIMD\cite{Skelton2013}	&	Crystal\cite{DelgadoGCT} 	\\ 
			\hline
			Cu 		&	2.65		&	2.57	&	2.79	&	2.61	\\
			Ge 		&	2.61		&	2.56	&	2.62	&	2.51	\\
			Te 		&	2.63		&	2.65	&	2.65	&	2.58	\\
			\hline 
		\end{tabular}
	\end{center}
\end{table}

\setlength{\tabcolsep}{4pt}
\begin{table}
	\caption{Partial and total coordination numbers of CGT in comparison with other studies. Partial coordinations $ij$ are given for $j$ atoms around $i$ centers. 
	}
	\label{tab:CNtable}
	\begin{center}
		\begin{tabular}{l|c|c|cc}  
			\hline							
			\multirow{ 2}{*}{elem.} & \multicolumn{2}{c|}{RMC results } & \multicolumn{2}{c}{references } \\
			\cline{2-5}
			&  present 	& limited  	& RMC\cite{JovariGCT} 	& AIMD\cite{Skelton2013} \\
			\hline
			CuGe 	&	0.73	&	-		&	-				&	0.62	\\
			CuCu 	&	2.38 	&	2.35	&	2.20$\pm0.4$	&	2.34	\\
			CuTe 	&	2.29	&	2.10	&	1.86$\pm0.3$	&	3.70	\\
			GeGe 	&	0.72	&	1.41	&	1.52$\pm0.4$	&	0.12	\\
			GeTe 	&	1.83	&	2.68	&	2.51$\pm0.5$	&	3.09	\\
			
			TeTe 	&	2.19	&	2.08	&	1.72$\pm0.3$	&	0.60	\\
			\hline
			$N$(Cu)	&	5.40	&	4.45	&	4.06$\pm0.6$	&	6.67	\\
			$N$(Ge)	&	4.02	&	4.09	&	4.03 \hspace{4ex}	&	4.47	\\  
			$N$(Te)	&	4.64	& 	4.41	&	4.10$\pm0.5$	&	4.18	\\
			$\langle N \rangle$	&	4.79 	&	4.37	&	4.08 \hspace{4ex} &	5.06	\\
			\hline
		\end{tabular}
	\end{center}
\end{table}       	

\section{Discussion}

\subsection{Coordination numbers and interatomic distances}
It is notable that - within the precision of the experimental methods - interatomic distances (see Table~\ref{tab:Disttable})
 in the amorphous phase of CGT do not differ largely from the  corresponding crystalline phase. 
 This indicates a remarkable similarity between the two phases, which is not found in other comparable phase-change materials: 
in GeSbTe, for example, the shortening of the Ge-Te bond was the basis for the proposed ``umbrella-flip'' model\cite{KolobovNatMat}.

We found that the obtained coordination numbers for the limited RMC approach in Table~\ref{tab:CNtable} are in agreement with the data by J\'{o}v\'{a}ri \textit{et al.}~\cite{JovariGCT} within the reported experimental uncertainties of $\pm$0.3-0.6.  
The effect of the AXS data in the present, full approach is mainly seen in the Ge environment, where a reduced number of Ge-Ge and Ge-Te bonds is found in favor of the Cu-Ge bonds. The existence of these bonds is difficult to judge from only XAFS and total scattering data, cf. ref.~\citen{JovariGCT}, but it is evident from the AXS data.  
Despite the disagreements between this model and any of the reference models, some consistent observations can be made: The structure of CGT is characterized by high average coordination numbers, especially around Cu; all coordination numbers are actually larger than in the corresponding crystal; also, a large number of homopolar bonds is found, especially Cu-Cu (for every model) and Te-Te (only in the experimentally obtained models) bonds.

\subsection{Bond angle distribution}
\begin{figure}
	\begin{center}   
		\includegraphics[width=0.75\linewidth]{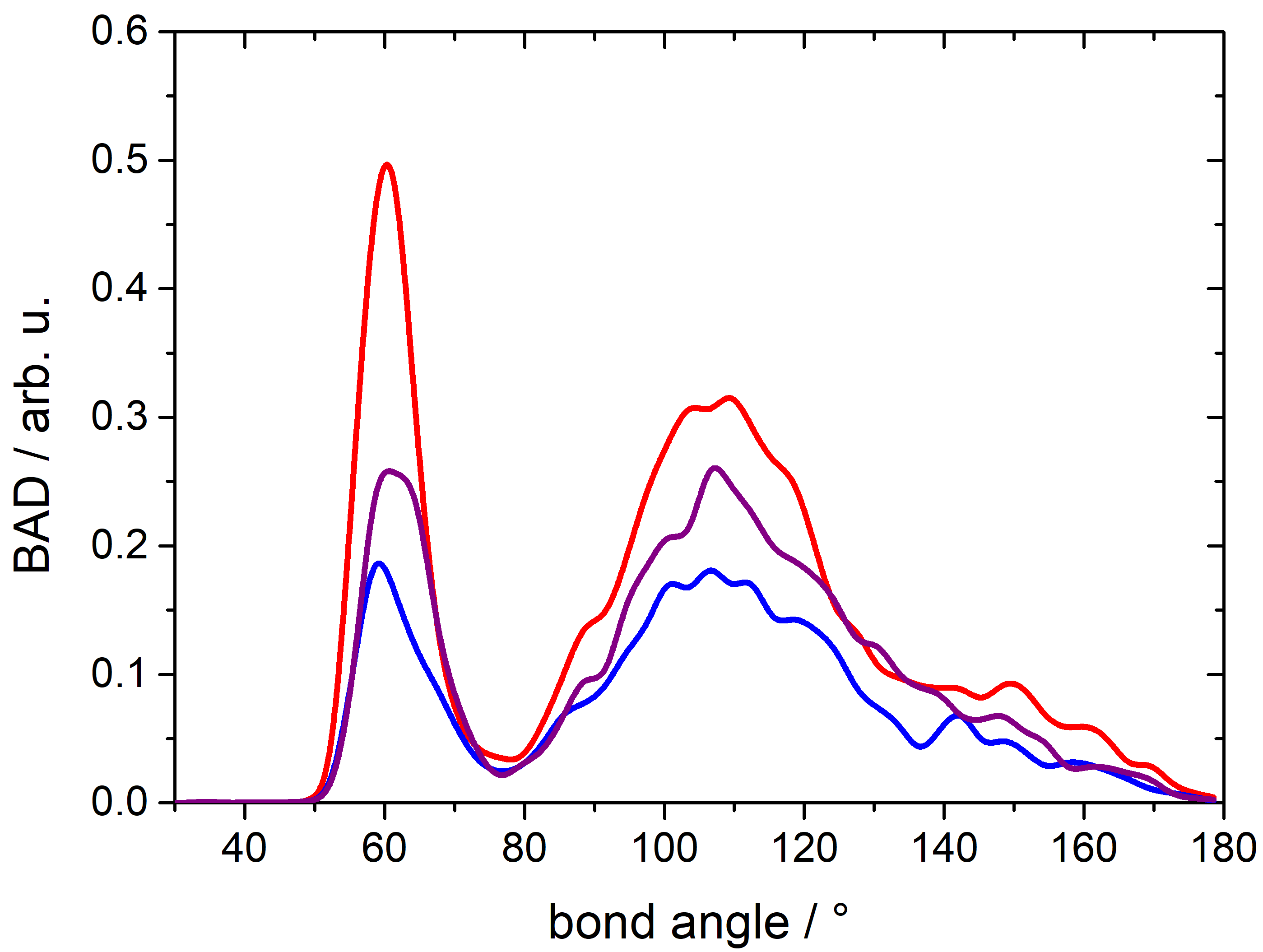}
		\caption{Bond angle distribution in a-CGT, around 
			Cu (red),
			Ge (blue) 	and 	
			Te atoms (purple).
		}
		\label{fig:BAD}
	\end{center}
\end{figure}

By including the AXS data, reliable information on structural features beyond the coordination numbers can be obtained. A detailed analysis of the present RMC model provides information on bond angle distributions (BAD) and ring statistics of the network. 
We calculated the BAD around the individual elements, shown in  Fig.~\ref{fig:BAD}.
In general, broad distributions around 109$^\circ$ are observed for all correlations, corresponding to a distorted tetrahedral coordination (109.5$^\circ$). This corresponds to the large number of 4-fold coordinated atoms, and shows a large similarity to the crystal structure, where only tetrahedral configurations are found, though with a much narrower distribution (104$^\circ$-114$^\circ$).
In addition, peaks around 60$^\circ$ are found and are mainly connected with Cu-related correlations. 
The results are consistent with theoretical studies.\cite{Skelton2013,Chen2015} 
The low number of 90$^\circ$ angles is a striking difference to GeSbTe-based PCMs,\cite{JovariJPCM, jovari2008, akola2009} and demonstrates that the amorphous phases of CGT and GeSbTe systems are dominated by very different structural motifs. 

\subsection{Ring statistics}
\begin{figure}
	\begin{center}		
		\includegraphics[width=0.7\linewidth]{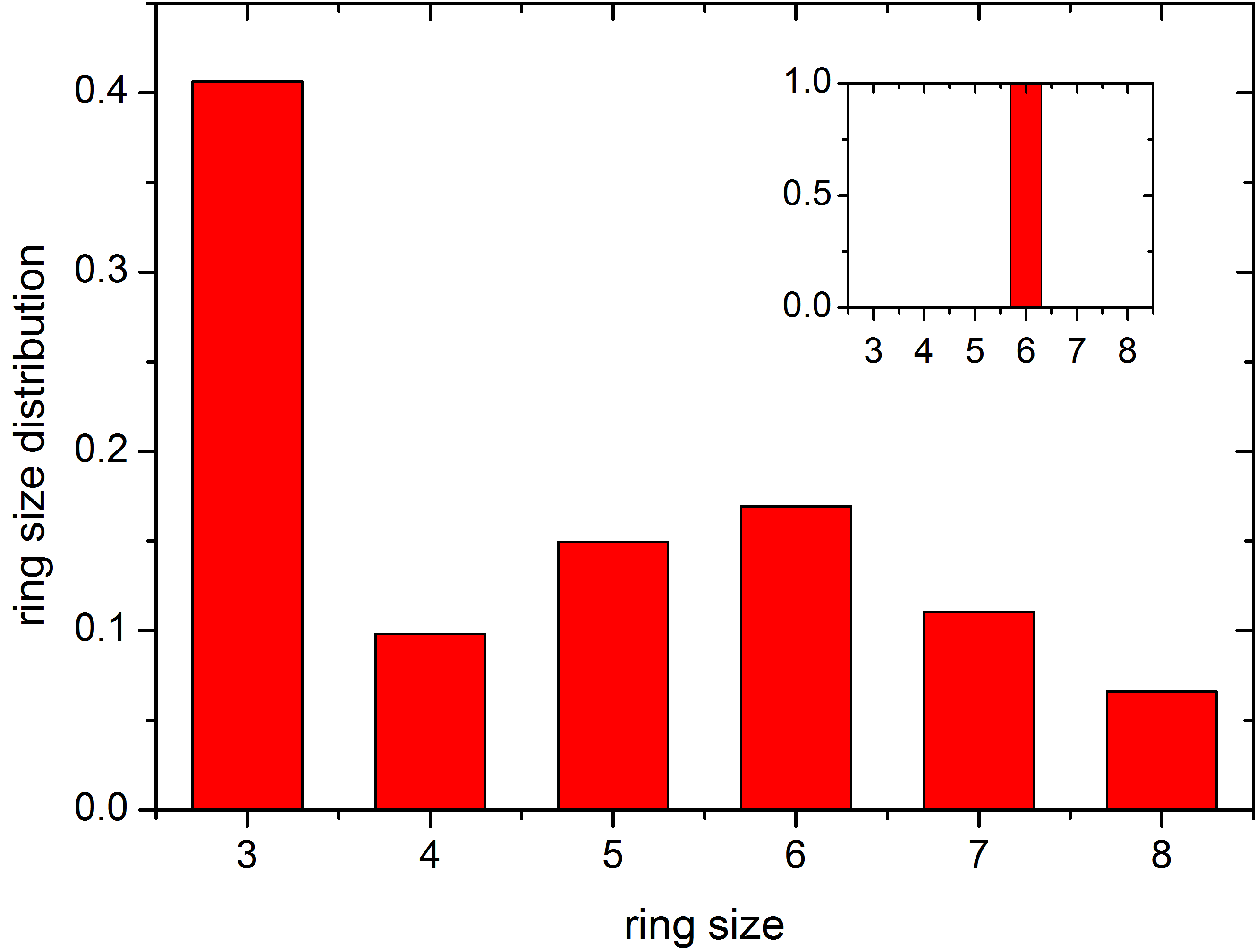}
		\caption{Ring statistics in a-CGT. Inset: c-CGT.}
		\label{fig:rings}
	\end{center}
\end{figure}

These features can be understood by considering the rings statistics, which were calculated using the R.I.N.G.S. program.\cite{RINGS} 
A ``ring'' is defined as a closed path of covalent bonds originating from and leading back to the same atom. For the ring statistics analysis, irreducible rings were searched in the amorphous network, i.e.\ closed paths that cannot be decomposed into smaller rings. The results are shown in Fig.~\ref{fig:rings}. 
A broad distribution of ring structures is found with a shallow maximum for 6-membered rings. This centering around the 6-rings shows a correspondence to the crystal structure, where  only 6-membered rings are found (inset in Fig.~\ref{fig:rings}). 
Furthermore, a large number of 3-fold rings is found, which corresponds to the peak around 60$^\circ$ in the BAD. A similar feature is observed even in an AIMD study.\cite{Chen2015}

\setlength{\tabcolsep}{6pt}
\begin{table*}
	\caption{Composition of the rings. The values denote the number of atoms of the respective element in the $n$-fold ring. The values in brackets indicate the relative increase from the ring composition expected by the concentration.}
	\label{tab:rings_composition}
	\begin{center}
		\begin{tabular}{c|cccccc}  
			\hline	
			element	&	$n$=3		&	4		&	5		&	6		&	7		&	8 	\\
			\hline 
			Cu		&	1.56	&	1.97	&	2.41	&	2.66	&	2.96	&	3.37	\\
			&	(+56\%)	&	(+48\%)	&	(+44\%)	&	(+33\%)	&	(+27\%)	&	(+26\%)		\\
			Ge		&	0.59	&	0.86	&	0.82	&	0.88	&	1.08	&	1.47	\\
			&	(+18\%)	&	(+28\%)	&	(-2\%)	&	(-12\%)	&	(-8\%)	&	(+10\%)		\\
			Te		&	0.85	&	1.18	&	1.78	&	2.46	&	2.96	&	3.16   \\
			&	(-43\%)	&	(-41\%)	&	(-29\%)	&	(-18\%)	&	(-15\%)	&	(-21\%)		\\
			\hline
		\end{tabular}
	\end{center}
\end{table*}

The significance of the large contribution of 3-fold rings was evaluated by an additional RMC run, in which  the formation of 3-rings was constrained by including a penalty term for 60$^\circ$ angles in the BAD.
Thereby, the number of 3-rings was reduced by 97\%; however, a significant increase was found in the goodness-of-fit values $R_w$ for the total $S(Q)$ by a factor of 1.40, and for the differential datasets $\Delta_kS(Q)$, especially for Ge (2.55), but also for Te (1.43) and to a smaller amount for Cu (1.14), indicating that 3-rings are an important component of the structure, and should be included in the modeling process. 
The $R_w$ values of the XAFS datasets increase only by a small amount (by a factor of 1.03-1.10). This indicates that information on bond angles and on the ring structure is not directly available from the XAFS data.  

GeSbTe-based PCMs show a markedly different ring distribution, where even-membered ring structures are supposed to be dominant.\cite{KoharaAPL, AkolaJones, akola2009} In GeSbTe, this structural feature is explained by a similarity to the crystal structure, where resonance bonding via $p$-orbitals (concomitant with 90$^\circ$ bond angles and 4-fold rings) plays an important role for the stability.\cite{shportko2008} 
The ring structures of CGT require a different explanation. For more details on the network, we analyzed the composition of the rings, shown in Table~\ref{tab:rings_composition}. The table displays the average number of atoms from a specific element in an $n$-fold ring (between $3<n<8$).  
The table also indicates the relative difference to the expected ring composition; for example, a 6-fold ring in Cu$_2$Ge$_1$Te$_3$ can be expected to consist of 2 Cu, 1 Ge and 3 Te atoms.   
In general, Cu is found in ring structures to a much larger degree than what would be expected from its relative concentration (+39\% on average). This finding agrees well with the high coordination number of the Cu atoms.  
For the most important ring sizes, typical building blocks of the rings structures are composed of:
\vspace{-1ex}
\begin{itemize}\setlength{\itemsep}{-0.2ex}
	\item 	3-fold rings: Cu$_2$Te,
	\item 	5-fold rings: Cu$_2$GeTe$_2$ 
	\item 	6-fold rings: Cu$_3$GeTe$_2$ and Cu$_2$GeTe$_3$
\end{itemize}	 
Except for Cu$_2$GeTe$_3$, which can be formed as an alternating ring structure (and is the only ring structure for the 
crystalline phase), these typical buildings blocks cannot be realized without the formation of ``wrong'' bonds of Cu-Ge or Cu-Cu.

\subsection{Model for the phase transition in CGT}
\begin{figure}
	\begin{center}   	
		\includegraphics[width=0.54\linewidth,valign=t]{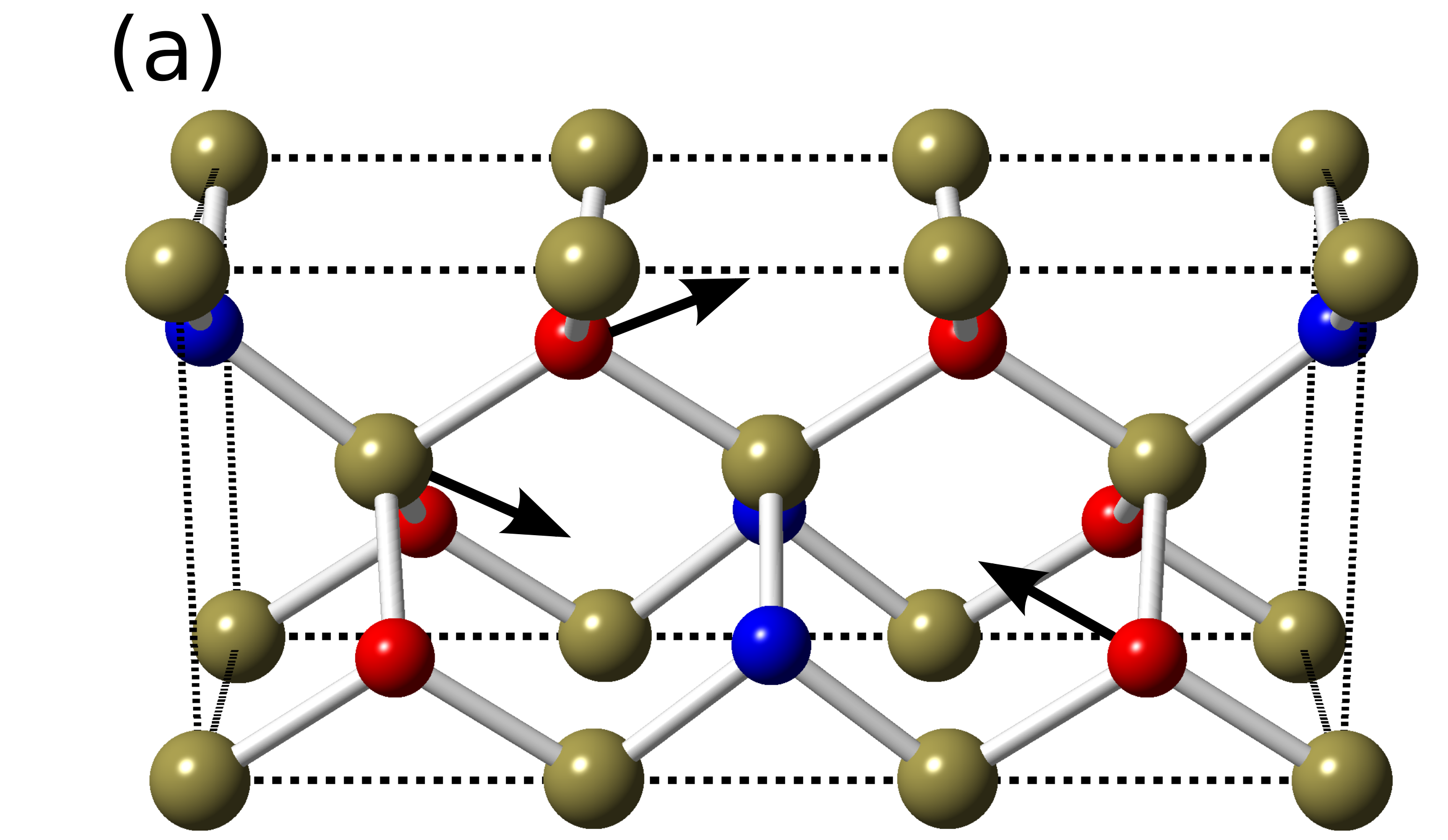}   
		\hspace{0.01\linewidth} 
		\includegraphics[width=0.42\linewidth,valign=t]{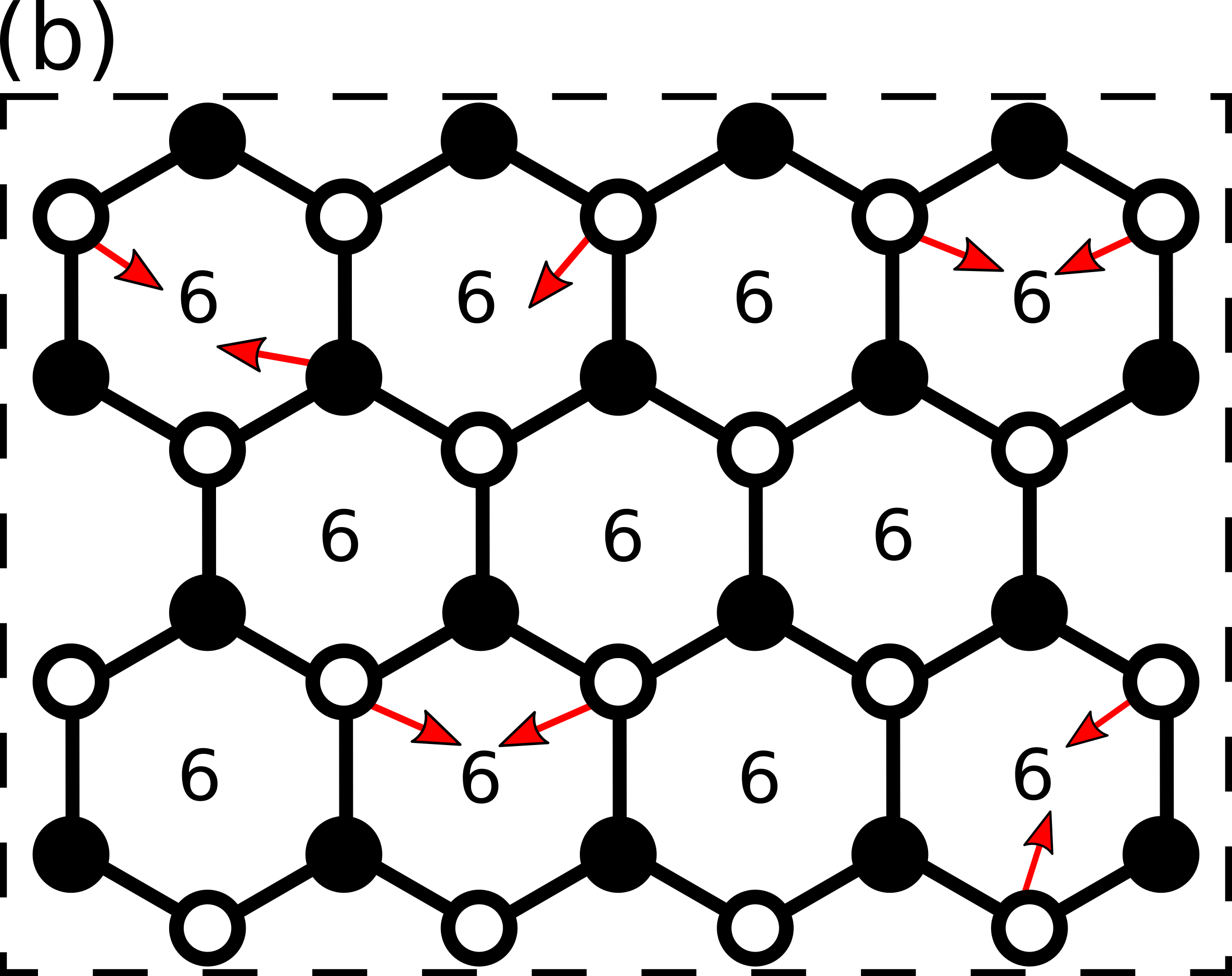} \\  
		\vspace{4ex}
		\includegraphics[width=0.54\linewidth,valign=t]{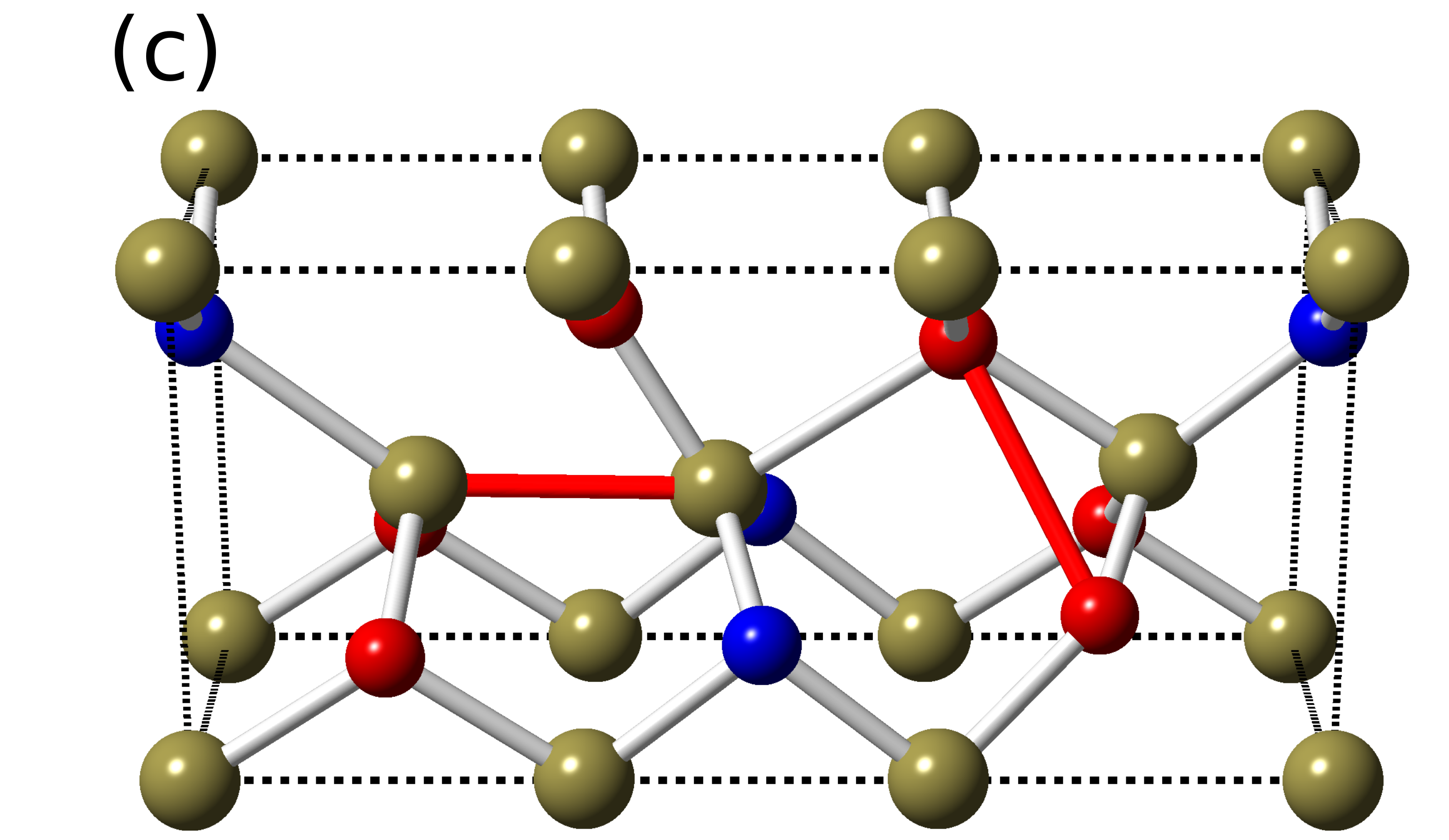}
		\hspace{0.01\linewidth} 
		\includegraphics[width=0.42\linewidth,valign=t]{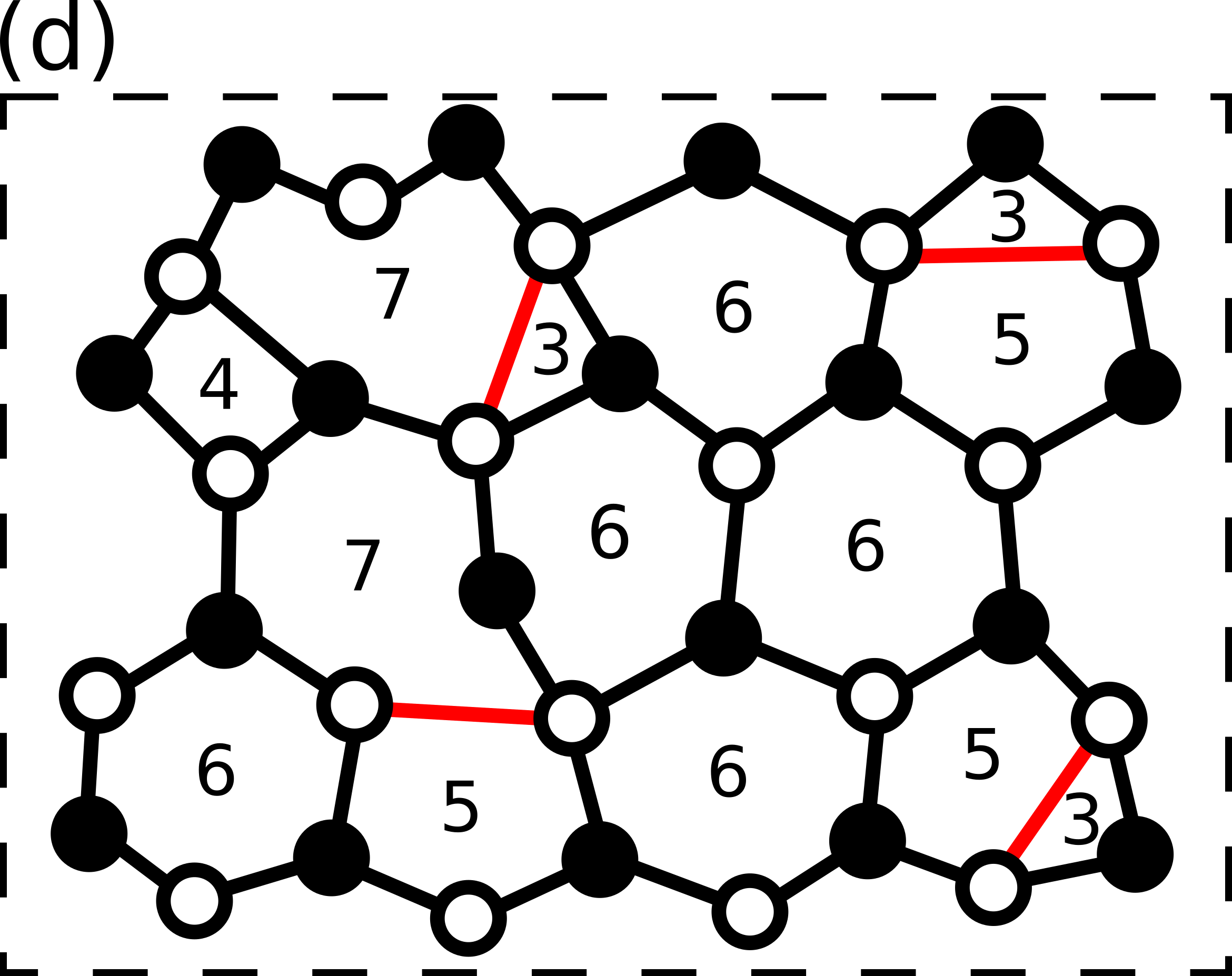}
		\caption{Model for the phase transition in CGT. The crystal structure is shown in (a) and a schematic representation of the ring structure in (b).
			During the phase transition, atoms in the crystal move towards the 6-ring centers, resulting in new ring structures and the formation of wrong bonds (red) in the amorphous phase, illustrated in (c) and (d).	Atomic movements resulting in new bonds are marked with arrows in (a) and (b). 	
			Colors denote
			Te (gold),
			Ge (blue) and 
			Cu (red) atoms. 
			Images of the structures were produced using VESTA.\cite{vesta} 
			In (b) and (d), filled circles denote Te and empty circles are Ge/Cu atoms. The numbers indicate the size  of the ring. 
		}
		\label{fig:model}
	\end{center}
\end{figure}

From the structural information described so far, it is possible to draw a detailed model of the amorphous structure, aimed to explain the fast structural phase transition in CGT. This model is illustrated in Fig.~\ref{fig:model}. 
Starting from the crystal structure, only small movements of the atoms are required to reach the amorphous state. Namely, there is movement of atoms (especially the Cu atoms) towards the centers of the 6-fold rings. These movements are illustrated in Fig.~\ref{fig:model}~(a) for one unit cell and in (b) schematically for the ring structure with arrow symbols. 
They lead to the increased coordination numbers of Te and Cu compared to the crystal and the increasing density of the amorphous phase.  
Note that the high mobility of the Cu atoms is also suggested as a key factor for the phase change process  
by a recent investigation of combined hard x-ray photoelectron spectroscopy and AIMD.\cite{Kobayashi2018}  
Concomitantly, this movement leads to the fragmentation of the 6-fold rings and the formation of smaller ring structures, shown in Fig.~\ref{fig:model}~(c) and~(d), in which ``wrong'' bonds of especially Cu-Cu and Te-Te are realized. 
The maximum at $n=6$ in the ring statistics  
reveals that a significant number of the 6-rings of the crystal structure are conserved, but become largely distorted, as indicated by the wide bond angle distribution in Fig.~\ref{fig:BAD}. 
This dominance of the 6-rings in the amorphous phase certainly contributes to the high speed of the phase transition. 
Finally, the redistribution of chemical bonds also leads to the formation of larger rings sizes with $n\geq7$.

Finally, we note that the focus so far has been the amorphization process. This formalism was chosen because it is straightforward to understand the structure of the amorphous phase as derived from the crystal. Technically, a fast crystallization, i.e.\ the reverse mechanism, is more important. This process can be understood as the reverse motion, i.e.\ in Fig.~\ref{fig:model} from (c,~d) to (a,~b), which in the same way requires only small atomic motions due to the similarities of the local structure.

\FloatBarrier
\section{Conclusion}

In conclusion, we present a model for the structure of the amorphous phase of Cu$_2$GeTe$_3$, based on the analysis of experimental data from AXS and XAFS, modeled by RMC. 
The extensive experimental approach represents a distinct improvement compared to previous experimental results.  
We confirmed the formation of smaller ring structures and a large number of homopolar bonds, in agreement with theoretical studies. The structural properties are used to  
draw a qualitative model of the phase-change process, in which atoms (especially Cu) move towards the centers of the 6-fold rings of the crystal, thereby forming new bonds and  resulting in a broader distribution of ring structures, but also preserving some structural motifs of the crystal, like the interatomic distances and the high coordination numbers.

\section{Acknowledgements}
The authors acknowledge partial financial support by the Japan Society for the Promotion of Science (JSPS) Grant-in-Aid for Scientific Research on Innovative Areas `3D Active-Site Science' (No.\ 26105006).
JRS also acknowledges financial support as Overseas researcher under a Postdoctoral Fellowship of JSPS (No.\ P16796). 
BP thanks the Fond der Chemischen Industrie for financial support.
The AXS experiments were performed at BM02 of the ESRF (Experimental nos.\ HC-2213 and HC-2534). 
The XAFS experiments were carried out at BL12C of the KEK-PF 
(Proposal nos.\ 2010G559 and 2012G522). 
We are indebted to L.~Pusztai (Wigner Research Centre for Physics, Hungary and Kumamoto University, Japan) for valuable discussions on the RMC data analysis, 
and to Y.~Saito and S.~Shindo (Tohoku University, Japan) for their help with the sample preparation.


\bibliography{GCT_arxiv}

\begin{thebibliography}{10}
\expandafter\ifx\csname url\endcsname\relax
  \def\url#1{\texttt{#1}}\fi
\expandafter\ifx\csname urlprefix\endcsname\relax\def\urlprefix{URL }\fi
\expandafter\ifx\csname href\endcsname\relax
  \def\href#1#2{#2} \def\path#1{#1}\fi

\bibitem{wuttig-nature2007}
M.~Wuttig, N.~Yamada, Phase-change materials for rewriteable data storage, Nat.
  Mater. 6 (2007) 824--832.

\bibitem{OvshinskyPRL}
S.~R. Ovshinsky, Reversible electrical switching phenomena in disordered
  structures, Phys. Rev. Lett. 21 (1968) 1450.

\bibitem{Sutou2012}
Y.~Sutou, T.~Kamada, M.~Sumiya, Y.~Saito, J.~Koike, Crystallization process and
  thermal stability of {Ge$_1$Cu$_2$Te$_3$} amorphous thin films for use as
  phase change materials, Acta Mater. 60 (2012) 872.

\bibitem{Saito2013}
Y.~Saito, Y.~Sutou, J.~Koike, Optical contrast and laser-induced phase
  transition in {Ge$_1$Cu$_2$Te$_3$} thin film, Appl. Phys. Lett. 102 (2013)
  051910.

\bibitem{Yamada1991}
N.~Yamada, E.~Ohno, K.~Nishiuchi, N.~Akahira, M.~Takao, Rapid-phase transitions
  of {GeTe-Sb$_2$Te$_3$} pseudobinary amorphous thin films for an optical disk
  memory, J. Appl. Phys. 69 (1991) 2849.

\bibitem{Saito2014}
Y.~Saito, Y.~Sutou, J.~Koike, Phase change characteristics in {GeTe-CuTe}
  pseudobinary alloy films, J. Phys. Chem. C 118 (2014) 26973--26980.

\bibitem{shportko2008}
K.~Shportko, S.~Kremers, M.~Woda, D.~Lencer, J.~Robertson, M.~Wuttig, Resonant
  bonding in crystalline phase-change materials, Nat. Mater. 7 (2008) 653--658.

\bibitem{DelgadoGCT}
G.~E. Delgado, A.~J. Mora, M.~Pirela, A.~Vel\'{a}squez-Vel\'{a}squez,
  M.~Villarreal, B.~J. Fern\'{a}ndez, Structural refinement of the ternary
  chalcogenide compound {Cu$_2$GeTe$_3$} by x-ray powder diffraction, Phys.
  Stat. Sol. A 201~(13) (2004) 2900--2904.

\bibitem{JovariGCT}
P.~J\'{o}v\'{a}ri, Y.~Sutou, I.~Kaban, Y.~Saito, J.~Koike, Fourfold coordinated
  te atoms in amorphous {GeCu$_2$Te$_3$} phase change material, Scr. Mater. 68
  (2013) 122.

\bibitem{Kamimura2016}
K.~Kamimura, S.~Hosokawa, N.~Happo, H.~Ikemoto, Y.~Sutou, S.~Shindo, Y.~Saito,
  J.~Koike, Xafs analysis on amorphous and crystalline new phase change
  material {GeCu$_2$Te$_3$}, J. Optoelectron. Adv. Mater. 18 (2016) 248.

\bibitem{Skelton2013}
J.~M. Skelton, K.~Kobayashi, Y.~Sutou, S.~R. Elliott, Origin of the unusual
  reflectance and density contrasts in the phase-change material
  {Cu$_2$GeTe$_3$}, Appl. Phys. Lett. 102 (2013) 224105.

\bibitem{Chen2015}
N.-K. Chen, X.-B. Li, X.-P. Wang, M.-J. Xia, S.-Y. Xie, H.-Y. Wang, Z.~Song,
  S.~Zhanga, H.-B. Suna, Origin of high thermal stability of amorphous
  {Ge$_1$Cu$_2$Te$_3$} alloy. a significant cu-bonding reconfiguration
  modulated by te lone-pair electrons for crystallization, Acta Mater. 90
  (2015) 88--93.

\bibitem{KoharaAPL}
S.~Kohara, K.~Kato, S.~Kimura, H.~Tanaka, T.~Usuki, K.~Suzuya, H.~Tanaka,
  Y.~Moritomo, T.~Matsunaga, N.~Yamada, Y.~Tanaka, H.~Suematsu, M.~Takata,
  Structural basis for the fast phase change of {Ge$_2$Sb$_2$Te$_5$}: Ring
  statistics analogy between the crystal and amorphous states, Appl. Phys.
  Lett. 89 (2006) 201910--1 -- 201910--3.

\bibitem{AkolaJones}
J.~Akola, R.~O. Jones, Structural phase transitions on the nanoscale the
  crucial pattern in the phase-change materials {Ge$_2$Sb$_2$Te$_5$} and
  {GeTe}, Phys. Rev. B 76 (2007) 235201--1--10.

\bibitem{akola2009}
J.~Akola, R.~O. Jones, Structure of amorphous {Ge$_8$Sb$_2$Te$_{11}$}
  {GeTe-Sb$_2$Te$_3$} alloys and optical storage, Phys. Rev. B 79 (2009)
  134118.

\bibitem{HegedusNM}
J.~Heged\"{u}s, S.~R. Elliott, Microscopic origin of the fast crystallization
  ability of {Ge-Sb-Te} phase-change memory materials, Nat. Mater. 7 (2008)
  399--405.

\bibitem{KolobovNatMat}
A.~V. Kolobov, P.~Fons, A.~I. Frenkel, A.~L. Ankudinov, J.~Tominaga, T.~Uruga,
  Understanding the phase-change mechanism of rewritable optical media, Nat.
  Mater. 3 (2004) 703--708.

\bibitem{hosokawa-GST}
S.~Hosokawa, W.-C. Pilgrim, A.~H\"ohle, D.~Szubrin, N.~Boudet, J.-F. B\'{e}rar,
  K.~Maruyama, Key experimental information on intermediate-range atomic
  structures in amorphous {Ge$_2$Sb$_2$Te$_5$} phase change material, J. Appl.
  Phys. 111 (2012) 083517.

\bibitem{Waseda1984}
Y.~Waseda, Novel Application of Anomalous (Resonance) X-Ray Scattering for
  Structural Characterization of Disordered Materials, {Berlin, Springer},
  1984.

\bibitem{KroghMoe}
J.~Krogh-Moe, A method for converting experimental x-ray intensities to an
  absolute scale, Acta Cryst. 9~(11) (1956) 951--953.

\bibitem{norman}
N.~Norman, The fourier transform method for normalizing intensities, Acta
  Cryst. 10 (1957) 370.

\bibitem{HosokawaPRB}
S.~Hosokawa, I.~Oh, M.~Sakurai, W.-C. Pilgrim, N.~Boudet, J.-F. B\'{e}rar,
  S.~Kohara, Anomalous x-ray scattering study of {Ge$_x$Se$_{1-x}$} glassy
  alloys across the stiffness transition composition, Phys. Rev. B 84 (2011)
  014201.

\bibitem{stellhorn-zpc}
J.~R. Stellhorn, S.~Hosokawa, W.-C. Pilgrim, Microscopic structure analysis in
  disordered materials using anomalous x-ray scattering, Z. Phys. Chem.
  228~(10-12) (2014) 1005--1031.

\bibitem{HosokawaZPC}
S.~Hosokawa, Y.~Wang, J.-F. B\'{e}rar, J.~Greif, W.-C. Pilgrim, K.~Murase,
  Anomalous x-ray scattering studies on glassy {Ge$_x$Se$_{1-x}$} across the
  stiffness threshold composition $x=0.20$, Z. Phys. Chem. 216 (2002)
  1219--1238.

\bibitem{HosokawaEPJST}
S.~Hosokawa, W.-C. Pilgrim, J.-F. B\'{e}rar, S.~Kohara, Anomalous x-ray
  scattering studies on semiconducting and metallic glasses, Eur. Phys. J.
  Special Topics 208 (2012) 291.

\bibitem{Gereben}
O.~Gereben, P.~J\'{o}v\'{a}ri, L.~Temleitner, L.~Pusztai, A new version of the
  {RMC++} reverse monte carlo programme, aimed at investigating the structure
  of covalent glasses, J. Optoelectron. Adv. Mater. 9 (2007) 3021--3027.

\bibitem{Gereben2012}
O.~Gereben, L.~Pusztai, {RMC\_POT}: A computer code for reverse monte carlo
  modeling the structure of disordered systems containing molecules of
  arbitrary complexity, J. Comput. Chem. 33 (2012) 2285--2291.

\bibitem{pyykko-covalent-radii}
P.~Pyykk\"{o}, Refitted tetrahedral covalent radii for solids, Phys. Rev. B 85
  (2012) 024115.

\bibitem{JovariJPCM}
P.~J\'{o}v\'{a}ri, I.~Kaban, J.~Steiner, B.~Beuneu, A.~Sch\"{o}p, A.~Webb,
  `{W}rong bonds' in sputtered amorphous {Ge$_2$Sb$_2$Te$_5$}, J. Phys.:
  Condens. Matter 19 (2007) 335212--1--9.

\bibitem{jovari2008}
P.~J\'{o}v\'{a}ri, I.~Kaban, J.~Steiner, B.~Beuneu, A.~Sch\"{o}p, A.~Webb,
  Local order in amorphous {Ge$_2$Sb$_2$Te$_5$} and {GeSb$_2$Te$_4$}, Phys.
  Rev. B 77~(3) (2008) 035202.

\bibitem{RINGS}
S.~L. Roux, P.~Jund, Ring statistics analysis of topological networks, Comp.
  Mater. Sci. 49 (2010) 70--83.

\bibitem{vesta}
K.~Momma, F.~Izumi, Vesta 3 for three-dimensional visualization of crystal,
  volumetric and morphology data, J. Appl. Cryst. 44 (2011) 1272--1276.

\bibitem{Kobayashi2018}
K.~Kobayashi, J.~M. Skelton, Y.~Saito, S.~Shindo, M.~Kobata, P.~Fons, A.~V.
  Kolobov, S.~Elliott, D.~Ando, Y.~Sutou, Understanding the fast phase-change
  mechanism of tetrahedrally bonded {Cu$_2$GeTe$_3$} comprehensive analyses of
  electronic structure and transport phenomena, Phys. Rev. B , 97 (2018)
  195105.

\end{thebibliography}

\end{document}